\documentclass[twocolumn]{revtex4}               
\usepackage{amsmath,amssymb}
\usepackage{latexsym}
\usepackage{epsfig}
\usepackage{color}
\begin{document}
\title{Kaniadakis holographic dark energy in Brans-Dicke cosmology}
\author{S. Ghaffari$^{1}$\footnote{sh.ghaffari@maragheh.ac.ir}}
\address{$^1$ Research Institute for Astronomy and Astrophysics of Maragha
(RIAAM), University of Maragheh, P.O. Box 55136-553, Maragheh,
Iran}

\begin{abstract}
By using the holographic hypothesis and Kaniadakis generalized entropy,
which is based on relativistic statistical theory and modified Boltzmann-Gibbs entrory,
we build Kaniadakis holographic dark energy (DE) model in the Brans-Dicke framework. We drive cosmological parameters of Kaniadakis holographic DE model, with IR cutoff as the Hubble horizon, in order to investigate its cosmological consequences.
Our study shows that, even in the absence of an interaction between the dark sectors of cosmos, the Kaniadakis holographic dark energy model with the Hubble radius as IR cutoff can explain the present accelerated phase of the universe expansion in the Brans-Dicke theory.
The stability of the model, using the squared of sound speed, has been checked and it is found that the model is unstable in non-interacting case and can be stable for some range of model parameters within the interacting case.
\end{abstract}
 \maketitle

\section{Introduction}

The possibility of justifying the current accelerated universe
using the energy density of quantum fields in vacuum leads to a
well-known model called holographic dark energy (HDE)
\cite{HDE3,HDE4,RevH}. In the original version, Bekenstein entropy
($S_{BH}$) and apparent horizon are employed as the entropy of a
gravitational system and IR cutoff, respectively \cite{HDE4}. Its
shortcomings motivate physicists to modify the model for example
by considering other IR cutoffs, or modified entropies
\cite{RevH,HDE5,HDE17,HDE01,HDE10,HDE12,stab}. Motivated by the
long range nature of gravity, recently, some new entropies have
been proposed as the alternatives of Bekenstein entropy which show
suitable results by themselves in cosmological setups
\cite{Tsallis,morad2016,morad2017,kum,Sayahian,Renyi,Tavayef,age,morad2020,maj}.

These entropies may also be justified by the quantum aspects of
gravity \cite{epl1,epl2,epjplus,barrow}, and additionally, the
corresponding HDE models provide acceptable description for the
current phase of the universe expansion even if the Hubble radius
is considered as the IR cutoff
\cite{Sayahian,Renyi,Tavayef,morad2020}. The latter is important
issue, because this horizon is thermodynamically the proper causal
boundary \cite{66,Renyi}, a property that motivates physicists to
deem the Hubble horizon as the IR cutoff. It was recently shown
that the Kaniadakis entropy of a black hole is obtained as
\cite{morad2020}

\begin{equation}\label{sk}
S_\kappa=\frac{1}{\kappa}\sinh(\kappa S_{BH}),
\end{equation}
where $ \kappa $ is a free parameter. Motivated by the
thermodynamic importance of the apparent horizon (Hubble horizon)
\cite{66}, authors in Ref.~\cite{morad2020} utilized this horizon as
the IR cutoff, and  proposed a new HDE model called KHDE (Kaniadakis
HDE) which seems to be able to justify the current accelerated expanding phase of the
universe. The applicability of KHDE in non-flat universe as well
as in the presence of other IR cutoffs have also been addressed in
Refs.~\cite{KHDEmorad,KHDE2}. In general, since the non-extensive
features of gravity, and also the properties of DE, are
not completely known, there is no restriction on the values of
$\kappa$ at present step, and more accurate observations as well as
other parts of physics may help us impose some constraints on its
values. Therefore, it is expectable to see different intervals for
$\kappa$ in meeting the observational requirement, depending on
the primary presumptions such as the IR cutoff
\cite{morad2020,KHDEmorad,KHDE2}.\\
On the other hand, the Brans-Dicke (BD) gravity is an
alternative to general relativity in which the gravitational
coupling $G$, is not a constant, and is replaced with the inverse
of a scalar field $(\phi)$ \cite{4not}. It is true that although
BD theory can explain accelerated expansion without resorting to DE
models, the obtained values for BD parameter ($\omega$) is lower
than the observational limit \cite{jcap,pav}. Therefore, there has been much attempt to solve this problem by introducing different
DE models in BD theory~\cite{Gong,Setare,Banerjee2,Banerjee,Xu,Jamil,Khodam}. Hence, according to dynamic behavior of the HDE, it is much appropriate to study it in the BD dynamic framework.
In this work, we would like to study the results of using
the KHDE model with the Hubble radius as the cut-off  in the BD
cosmology. In the next section, we extract the KHDE
density in BD gravity and investigate its cosmological parameters
for the non-interacting case. In section III, we consider interaction KHDE model and investigate its cosmological evolution. Section IV is devoted to stability of model and our conclusions are drawn in section V.

\section{ Non-interacting Kaniadakis holographic dark energy in the Brans-Dicke cosmology}
We consider a homogeneous and isotropic FRW universe described by the line element
\begin{equation}
{\rm d}s^2=-{\rm d}t^2+a^2(t)\left(\frac{{\rm d}r^2}{1-kr^2}+r^2{\rm
d}\Omega^2\right),\label{metric}
\end{equation}
where $k=0,1,-1$ represent a flat, closed and open maximally
symmetric space, respectively. The Brans-Dicke field equations can be written as \cite{Banerjee}

\begin{equation}\label{Friedeq01}
\frac{3}{4\omega}\phi^2\Big(H^2+\frac{k}{a^2}\Big)-\frac{\dot{\phi}^2}{2}+\frac{3}{2\omega}H\dot{\phi}\phi=\rho_M+\rho_\Lambda,
\end{equation}
\begin{equation}\label{Friedeq02}
\frac{-\phi^2}{4\omega}\Big(\frac{2\ddot{a}}{a}+H^2+\frac{k}{a^2}\Big)-\frac{1}{\omega}H\dot{\phi}\phi
-\frac{1}{2\omega}\ddot{\phi}\phi-\frac{\dot{\phi}^2}{2}\Big(1+\frac{1}{\omega}\Big)=p_\Lambda,
\end{equation}
\begin{equation}\label{motiom eq}
\ddot{\phi}+3H\dot{\phi}-\frac{3}{2\omega}\Big(\frac{\ddot{a}}{a}+H^2+\frac{k}{a^2}\Big)\phi=0.
\end{equation}
where $\phi^2=\omega/(2\pi G_{\rm eff})$, $G_{\rm eff}$ is the effective gravitational constant \cite{4not}, $H=\dot{a}/a $ is the Hubble parameter and $ \rho_m $, $ \rho_\Lambda $ and $ p_\Lambda $ are the pressureless dark matter density,
dark energy density and pressure of DE, respectively.
Following \cite{Banerjee2}, we assume that the BD field $\phi$
is proportional to scalar field as $(\phi \propto a^n) $, then we get

\begin{equation}\label{phidot}
\dot{\phi}=nH\phi,
\end{equation}
and hence
\begin{equation}\label{phiddot}
\ddot{\phi}=n^2H^2\phi+n\dot{H}\phi,
\end{equation}
where a dot denotes derivative with respect to time.
According to the Kaniadakis generalized entropy, that it is independent of gravitational theory, and holographic principle
to introduce a KHDE model in BD gravity, one arrives at the following relation for the Kanadakis holographic energy density(KHDE)
in which the apparent horizon in the flat Universe is considered as the IR cut-off $ (L=H^{-1}) $
\begin{equation}\label{rho1}
\rho_D=\frac{3c^2\phi^2H^4}{4\omega\kappa}\sinh\left(\frac{2\kappa \pi^2\phi^2}{\omega H^2}\right).
\end{equation}
Here, $  c^2 $ is an unknown constant~\cite{HDE3}. 
It is clear that in the limit $ \kappa\rightarrow 0 $ and $ G_{\rm eff} \rightarrow G $,
the energy density of HDE in standard cosmology is restored.
Also, for the limiting case $ G_{\rm eff} \rightarrow G $, the above equation yields 
the KHDE density in the standard Einstein gravity \cite{KHDEmorad}.
Defining the critical density as $ \rho_{cr}=\frac{3\phi^2H^2}{4\omega} $, 
the dimensionless density parameters are presented by
\begin{eqnarray}\label{Omega}
\Omega_m=\frac{\rho_m}{\rho_{cr}}&=&\frac{4\omega\rho_m}{3\phi^2H^2},\nonumber\\
\Omega_D=\frac{\rho_D}{\rho_{cr}}&=& \frac{4\omega\rho_D}{3\phi^2H^2},\nonumber\\
\Omega_\phi=\frac{\rho_\phi}{\rho_{cr}}&=& 2n\Big(\frac{n\omega}{3}-1\Big).
\end{eqnarray}
Using the above definition, we can write Eq. (\ref{Friedeq01}) as follows
\begin{equation}\label{Friedeq03}
\Omega_m+\Omega_D=1+\Omega_\phi.
\end{equation}
Using the second line of definitions (\ref{Omega}) along with Eq. (\ref{rho1}), the dimensionless
density parameter of KHDE is obtained as
\begin{equation}\label{Omega1}
\Omega_D=\frac{c^2H^2}{\kappa}\sinh\left(\frac{2\kappa\pi^2\phi^2}{\omega H^2}\right).
\end{equation}
Here, we also assume that two dark sectors of the universe do not interact with each other, i.e., there is no energy exchange between these cosmic sectors, then the energy conservation equations are given as follow
\begin{equation}
\dot{\rho}_D+3H(1+\omega_D)\rho_D=0,\label{ConserveDE}
\end{equation}
and
\begin{equation}
\dot{\rho}_m+3H\rho_m=0,\label{ConserveCDM}
\end{equation}
where $\omega_D=\frac{p_D}{\rho_D}$ denotes the equation of state (EoS) parameter of DE.
Taking the time derivative of Eq. (\ref{rho1}), we have
\begin{eqnarray}\label{rhodot}
\dot{\rho}_D&=&2H\rho_D\left(n+\frac{\dot{2H}}{H^2}\right)\nonumber\\&+&\frac{3\pi c^2\phi^2H^3}{2\omega}
\cosh\left(\frac{2\kappa\pi^2\phi^2}{\omega H^2}\right)\left(n-\frac{\dot{H}}{H^2}\right).
\end{eqnarray}
Taking the time derivative of Eq. (\ref{Friedeq01}) and combining the result with Eqs.  (\ref{phidot}),
(\ref{phiddot}), (\ref{ConserveCDM}) and (\ref{rhodot}), one can easily get
\begin{eqnarray}\label{Hdot1}
\frac{\dot{H}}{H^2}&=&\Big[-\frac{9\Omega_m^0H_0^2\phi_0^2}{4\omega \phi^2 H^2}+(2\omega n^2-6n-3)\frac{n}{2\omega}\nonumber\\\!\!\!+&&\!\!\!\!\!\!\!\!\!\!\!\!
\frac{3nc^2\pi^2\phi^2}{\omega^2}\cosh\left(\frac{2\kappa\pi^2\phi^2}{\omega H^2}\right)+
\frac{3nc^2H^2}{2\omega\kappa}\sinh\left(\frac{2\kappa\pi^2\phi^2}{\omega H^2}\right)\Big]\nonumber\\\!\!\!\times&&\!\!\!\!\!\!\!\!\!\!\!\!
 \Bigg[\frac{3c^2\pi^2\phi^2}{\omega^2}\cosh\left(\frac{2\kappa\pi^2\phi^2}{\omega H^2}\right)
 -\frac{3c^2H^2}{\omega\kappa}\sinh\left(\frac{2\kappa\pi^2\phi^2}{\omega H^2}\right)\Bigg]^{-1},\nonumber\\
\end{eqnarray}
 where $\Omega_m^0$, $H_0$ and $\phi_0$ are present values of matter density parameter, hubble parameter and BD field, respectively. Combining Eqs. (\ref{ConserveDE}) and (\ref{rhodot}), one can obtain the EoS parameter for the KHDE model in BD gravity as
\begin{eqnarray}\label{EoS1}
&&\omega_D=-1-\frac{2n}{3}-\frac{4\kappa\pi^2\phi^2}{3\omega H^2}\coth\left(\frac{2\kappa\pi^2\phi^2n}{\omega H^2}\right)\\&&\nonumber
-\frac{4}{3}\Big[1-\frac{4\kappa\pi^2\phi^2}{3\omega H^2}\coth\left(\frac{2\kappa\pi^2\phi^2}{\omega H^2}\right)\Big]\frac{\dot{H}}{H^2}.
\end{eqnarray}
One can also find out that, in the limiting case $ \kappa\rightarrow 0 $,
this equation is reduced to the EoS parameter for the original HDE model in BD gravity~\cite{Ghaffari1}.
It is also clear that the EoS parameter for KHDE in standard cosmology is recovered at
the appropriate limit $ n\rightarrow0 $ and $ (\omega\rightarrow 0) $ \cite{KHDEmorad}.

Also, using the deceleration parameter relation $ q=-1-\frac{\dot{H}}{H^2} $ and Eq. (\ref{Hdot1}), one gets
\begin{eqnarray}\label{q1}
&&q=-1-\Big[-\frac{9\Omega_m^0H_0^2}{4\omega H^2}+(2\omega n^2-6n-3)\frac{n}{2\omega}\nonumber\\&&+
\frac{3nc^2\pi^2\phi^2}{\omega^2}\cosh\left(\frac{2\kappa\pi^2\phi^2}{\omega H^2}\right)
\!+\!\frac{3nc^2H^2}{2\omega\kappa}\sinh\left(\frac{2\kappa\pi^2\phi^2}{\omega H^2}\right)\Big]\nonumber\\&&\times
\Bigg[\frac{3c^2\pi^2\phi^2}{\omega^2}\cosh\left(\frac{2\kappa\pi^2\phi^2}{\omega H^2}\right)\!-\!\frac{3c^2H^2}{\omega\kappa}\sinh\left(\frac{2\kappa\pi^2\phi^2}{\omega H^2}\right)\Bigg]^{-1}.\nonumber\\
\end{eqnarray}
One can also examine the fate of the universe filled with DM and KHDE components.
To this end, we should consider the effective EoS parameter
\begin{equation}
\omega_{\rm eff}=\frac{P}{\rho}=\frac{p_{DE}}{\rho_M+\rho_D}.
\end{equation} 
The behavior of the $ \omega_{\rm eff} $ is plotted in Figs. \ref{figweff1} and \ref{figqQ}.
\subsection{Cosmological Evolution}
\begin{figure}[htp]
\begin{center}
\includegraphics[width=8cm]{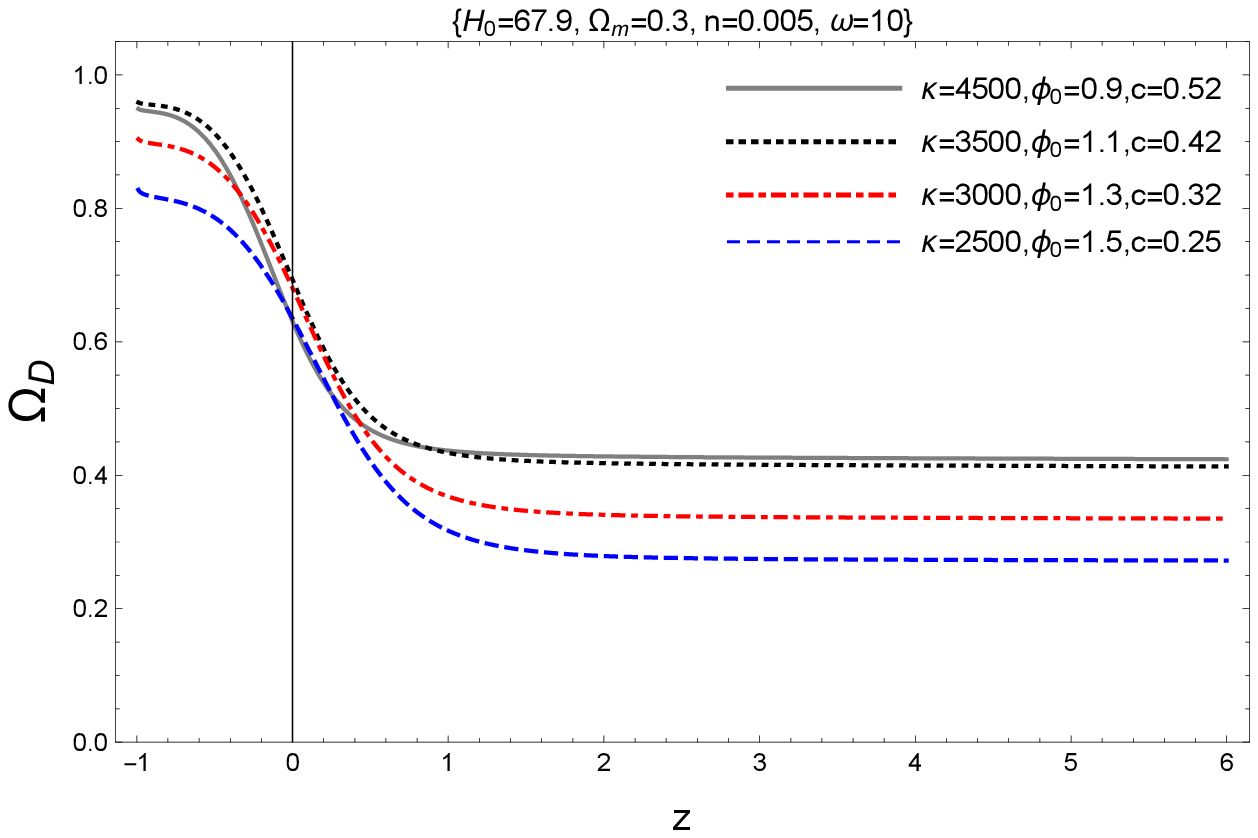}
\includegraphics[width=8cm]{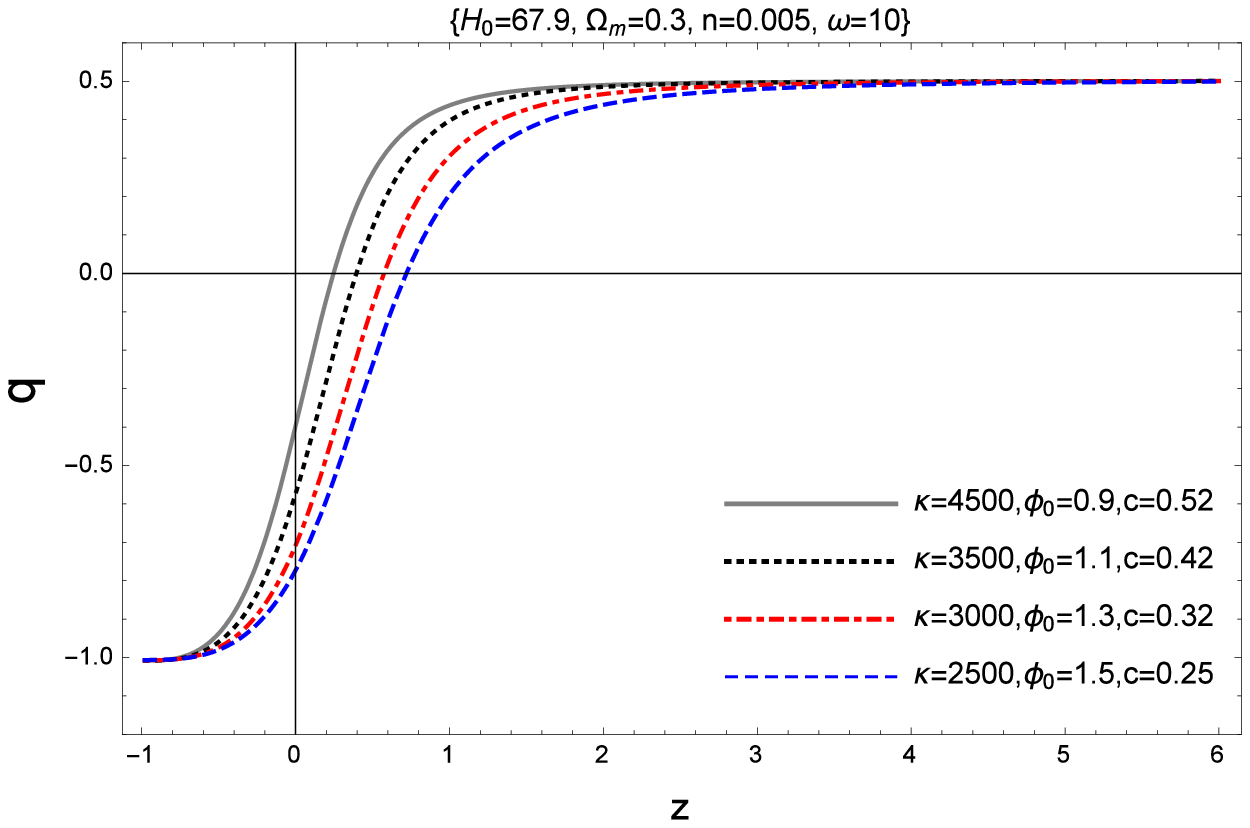}
\includegraphics[width=8cm]{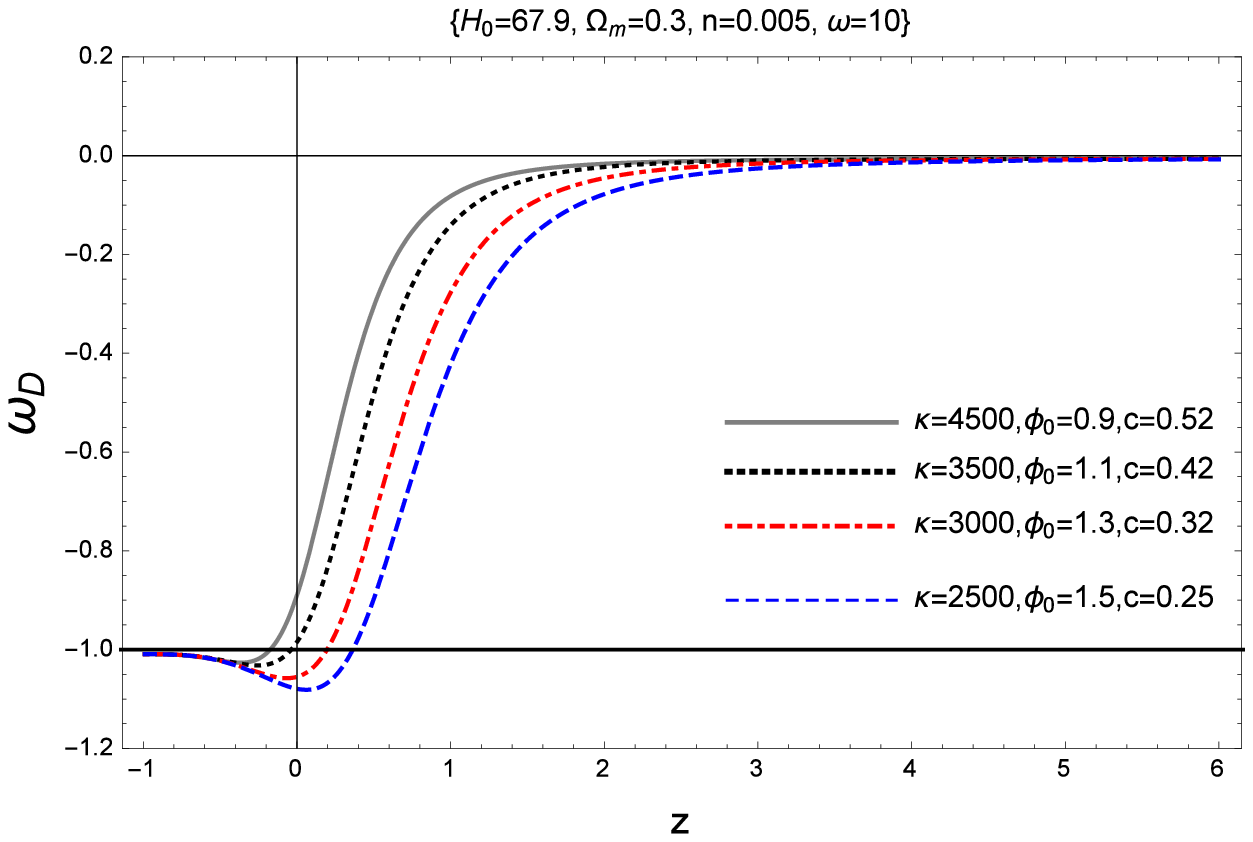}
 \caption{The evolution of the $ \omega_D(z) $, $ q(z) $ and  $ \omega_D(z) $ parameters for Non-interacting KHDE in BD gravity,
 in which fix parameters $H_0=67.9, \Omega_m=0.3, n=0.005, \omega=10 $ and various values of  $ \kappa,\phi_0, c $ are adopted.}
\label{fig1}
\end{center}
\end{figure}

In what follows, we investigate the cosmological behavior of a universe filled with KHDE and DM.
As it is apparent form Eqs. (\ref{Omega1}), (\ref{EoS1}) and (\ref{q1}), the cosmological parameters such as  
dimensionless density, EoS and deceleration parameters completely depend on the behavior of Hubble parameter 
function $ H(z) $ and its differential equation with number of constant parameters.
 As differential equation (\ref{Hdot1})  cannot be solved analytically, we proceed with solving  it numerically
for fixed  values $H_0=67.9 $ \cite{HubbleOb}, $\Omega_m=0.3 $ \cite{OmegamOb}, $ n=0.005$, $\omega=10 $ at  the present time $ (z=0) $.\\
In the upper panel of Fig. (\ref{fig1}) we have sketched the behavior of dimensionless density parameter $ \Omega_D $ in terms of redshift. 
From this figure we clearly see that at the early times of the universe ($ z\rightarrow \infty $), we have $ \Omega_D\rightarrow 0.2 $ in full agreement with Eq. (\ref{Friedeq03}), while at the late-time ($ z\rightarrow-1 $) the DE dominates ($ \Omega_D\rightarrow1 $). It is clear that for decreasing values of  $ c $ and $ \kappa $ parameters $ \Omega_D $ become smaller at the late time, which means that more energy is transferred to the matter component.\\
In the middle panel of Fig. (\ref{fig1}) we have plotted the evolution of the deceleration parameter $ q(z) $ versus redshift $ z $
where we have considered the initial condition as mentioned above and various values of $ c, \kappa $ and $ \phi_0 $
in order to study the effects of the model parameters. As we can see, the KHDE model with the Hubble cutoff in BD gravity can describe the current accelerated expansion of the universe, even in the absence of an interaction between the two dark sectors of cosmos unlike the standard HDE in the BD gravity~\cite{Xu}. The transition from deceleration to acceleration is realized around $ 0.25<z<0.8$ which is quite  in agreement with cosmological observations \cite{Daly,Komatsu,Salvatelli}. 
Also, we observe, the behavior of the deceleration parameter depends on the values of $c, \kappa $ and $ \phi_0 $. For decreasing model parameters $ c $ and $ \kappa $, the transition point moves to older universe and also the value of deceleration parameter at the present time  $ q(z=0)=q_0 $ decreases.\\ 
Finally, in the lower panel, we depict $ \omega_D(z) $ for various values of $ \kappa,\phi_0, c $ parameters.
As is clear, there is  a transition from the quintessence regime to the phantom regime at the certain values of redshift.
This value of the redshift depends on the  $ c $ and $ \kappa $  parameters of KHDE model, so that,
with increasing values of  $ c $ and $ \kappa $ this transition will occur in the near present time (smaller redshift) and also 
the EoS parameter at the present time $ (\omega_D(z=0)) $ will be smaller.

\begin{figure}[htp]
\begin{center}
\includegraphics[width=9cm]{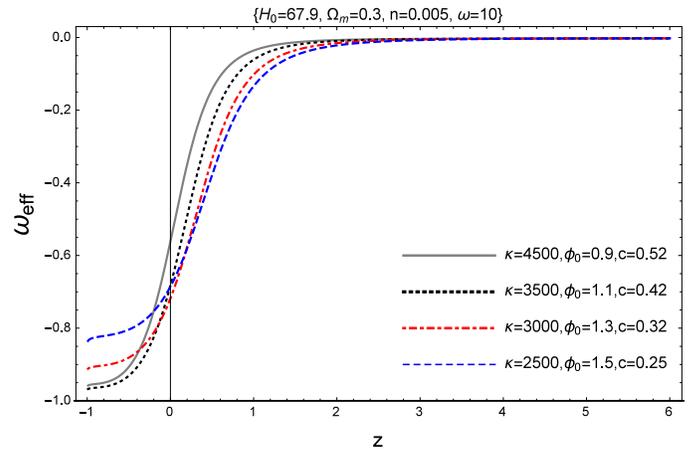}	
\caption{The evolution of the $ \omega_{\rm eff}(z) $ for non-interacting KHDE in BD gravity,
in which fix parameters $H_0=67.9, \Omega_m=0.3, n=0.005, \omega=10 $ and various values of  $ \kappa,\phi_0, c $ are adopted.}
\label{figweff1}
\end{center}
\end{figure}

We have plotted the evolution of $ \omega_{\rm eff}(z) $ for KHDE in BD gravity in Fig. \ref{figweff1}.
As it is obvious in this figure, at the early time we have the $ \omega_{\rm eff}\rightarrow 0 $, which shows a pressureless DM
dominated universe, while at the late time we have $\omega_{\rm eff}\rightarrow-1$, which means ending of the universe in a Big-Rip singularity.

\section{Interacting KHDE model}
In this section, we focus on the interaction between KHDE model and DM.
For the FRW universe filled with DE and DM interacting with each other, the
conservation equations are given by
\begin{equation}
\dot{\rho}_D+3H(1+\omega_D)\rho_D=-Q,\label{QConserveDE}
\end{equation}
\begin{equation}
\dot{\rho}_m+3H\rho_m=Q,\label{QConserveCDM}
\end{equation}
where $Q$ represents the interaction term, and we assume that it has the
form $Q=H(\alpha\rho_m+\beta\rho_D)$~\cite{pavonQ}, in which $ \alpha $ and $\beta $ are coupling  constants.
Taking the time derivative of Eq.~(\ref{Friedeq01}) along with using Eqs.~(\ref{phidot}),
(\ref{phiddot}),~(\ref{rhodot}) and~(\ref{QConserveCDM}), we get

\begin{eqnarray}\label{Hdot2}
&&\frac{\dot{H}}{H^2}=\Bigg[-\frac{3\Omega_m^0H_0^2H_0^2}{4\phi^2H^2}+(2\omega n^2-6n-3)\frac{n}{6}+\frac{\alpha\Omega_m}{4}\nonumber\\&+&\!\!\!
\frac{nc^2\pi^2\phi^2}{\omega}\cosh\left(\frac{2\kappa\pi^2\phi^2}{\omega H^2}\right)+\frac{c^2H^2(n+\frac{\beta}{2})}{2\kappa}\sinh\left(\frac{2\kappa\pi^2\phi^2}{\omega H^2}\right)\Bigg]\nonumber\\&\times&\!\!\!\nonumber
\Bigg[\frac{c^2\pi^2\phi^2}{\omega}\cosh\left(\frac{2\kappa\pi^2\phi^2}{\omega H^2}\right)-\frac{c^2H^2}{\kappa}\sinh\left(\frac{2\kappa\pi^2\phi^2}{\omega H^2}\right)\\&+&\!\!\!n-\frac{\omega n^2}{3}+\frac{1}{2}\Bigg]^{-1}.
 \end{eqnarray}
In the absence of interaction term $ (\alpha=\beta=0) $, Eq. (\ref{Hdot2}) is reduced to
its respective relation in the previous section. 

Combining Eqs.~(\ref{QConserveDE}) and~(\ref{rhodot}), one can obtain the following expression for EoS parameter
\begin{eqnarray}\label{EoS2}
\omega_D&=&-1+\frac{1}{3}(\alpha-\beta-2n)\\&-&\nonumber\frac{4\kappa\pi^2\phi^2}{3\omega H^2}\coth\left(\frac{2\kappa\pi^2\phi^2n}{\omega H^2}\right)\left(n-\frac{\dot{H}}{H^2}\right)\\&+&\nonumber
\frac{\kappa\alpha(2\omega n^2-2n-1)}{9c^2H^2\sinh\left(\frac{2\kappa\pi^2\phi^2n}{\omega H^2}\right)}.
\end{eqnarray}
\subsection{Cosmological Evolution}
\begin{figure}[htp]
\begin{center}
\includegraphics[width=8cm]{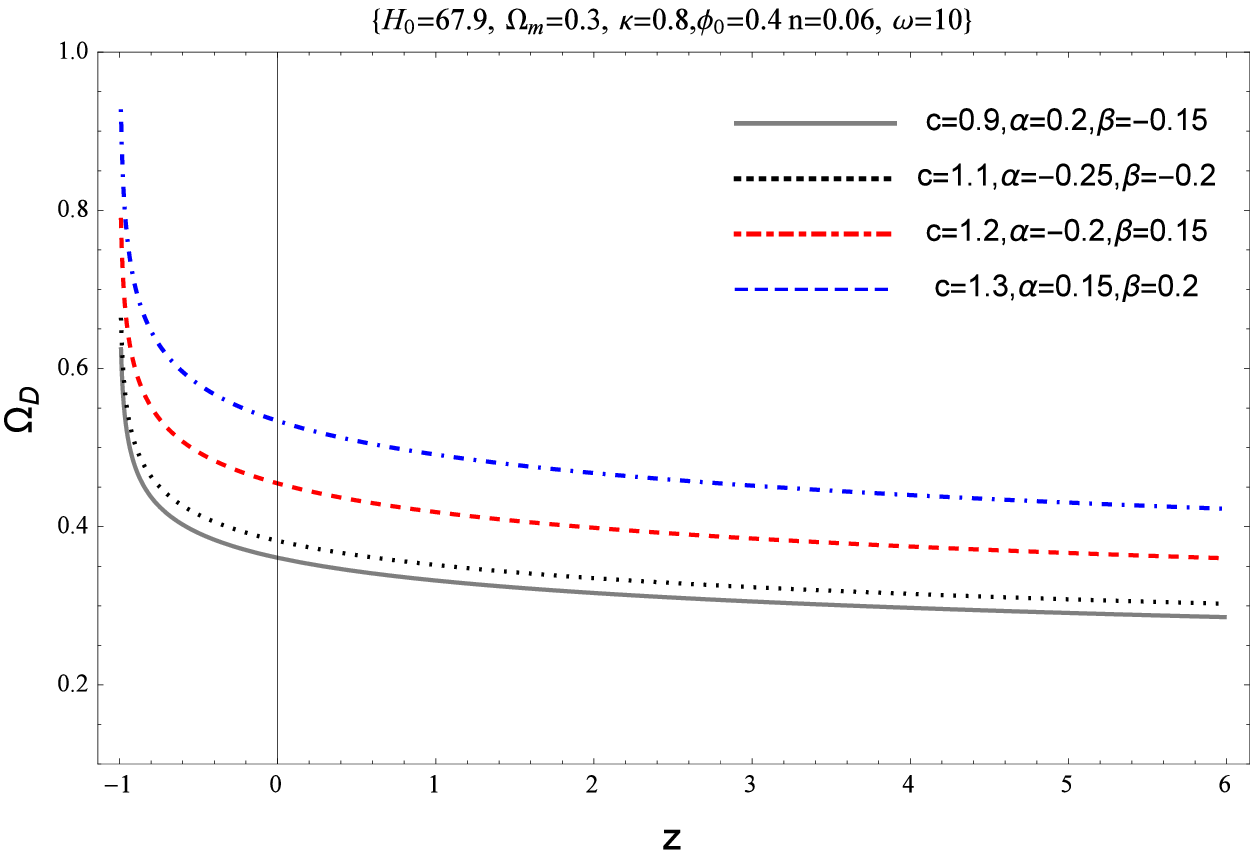}
\includegraphics[width=8cm]{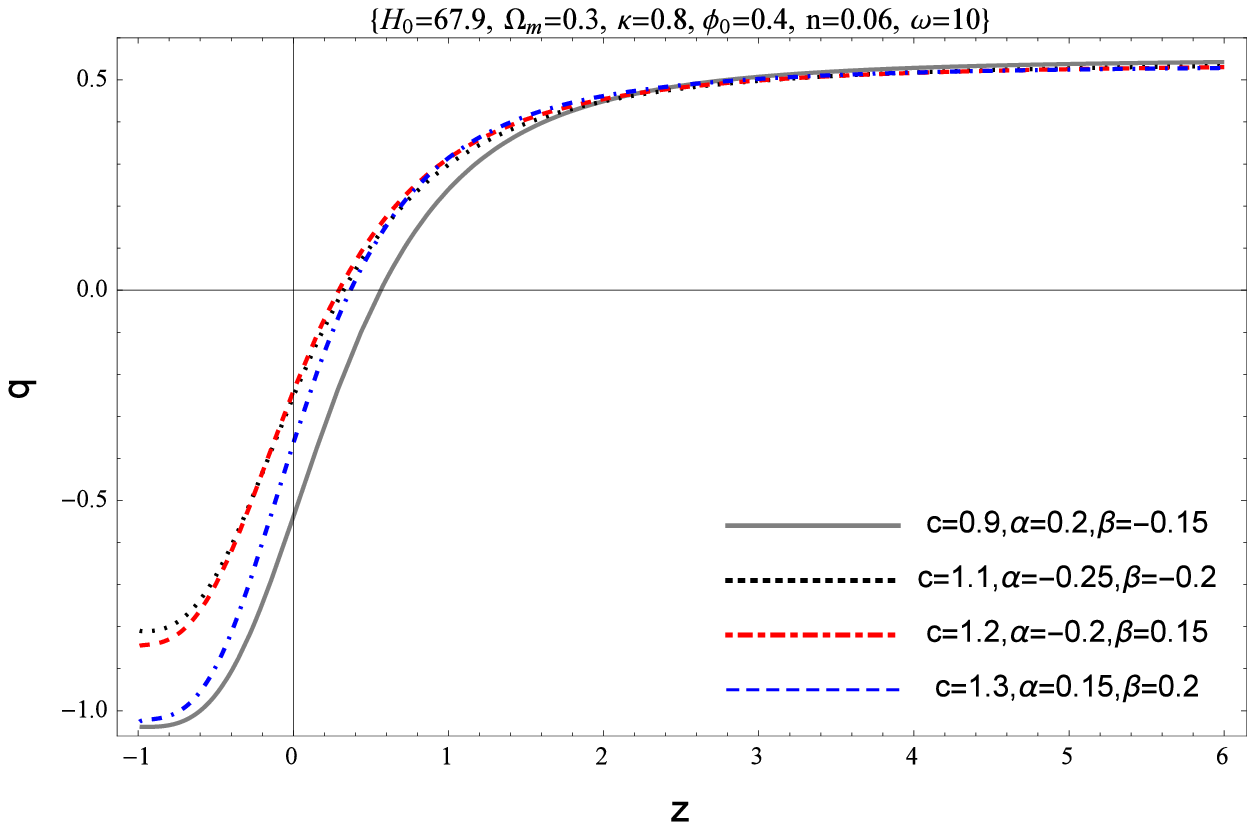}
\includegraphics[width=8cm]{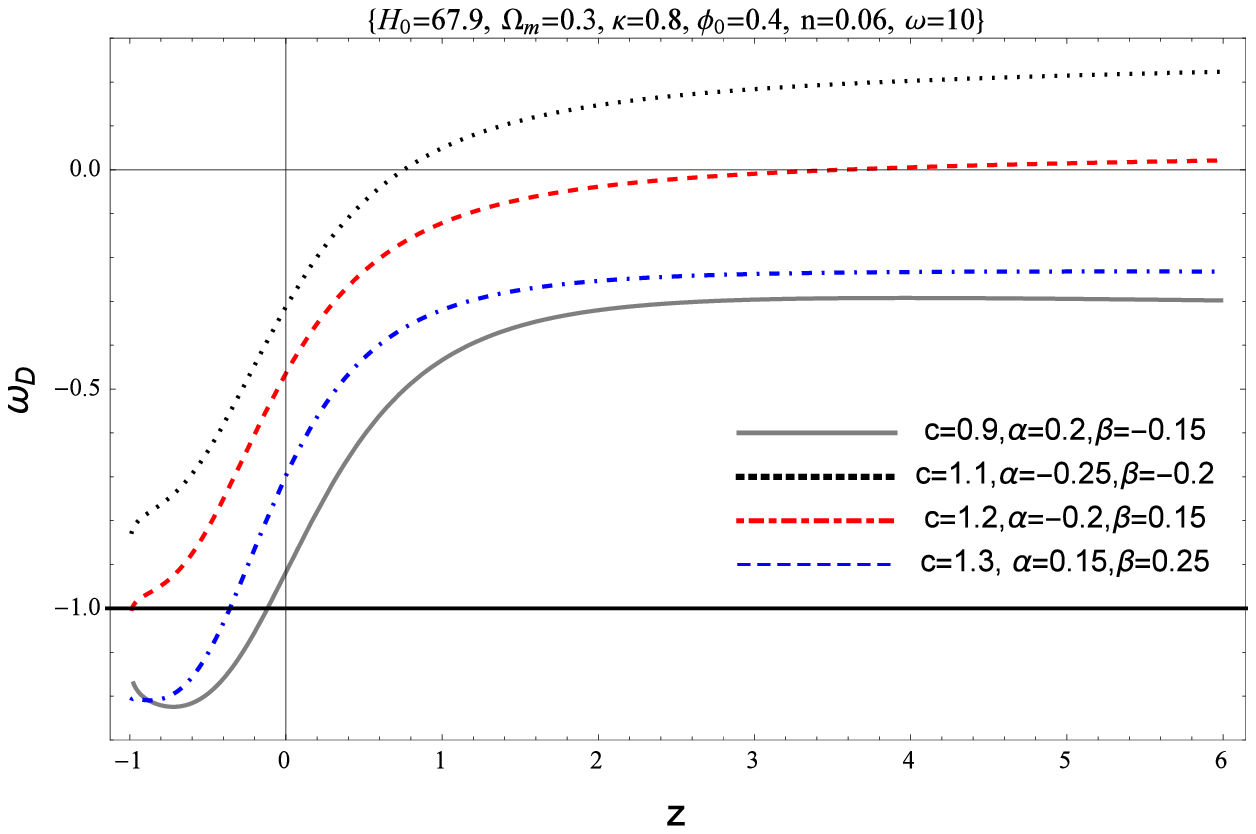}
\caption{The evolution of the $ \Omega_D(z) $, $ q $ for interacting KHDE for different para. 
We have taken the $ H_0=67.9, \Omega_m=0.3, \kappa=0.8, \phi_0=0.4, n=0.06 $ and $ \omega=10 $ 
as the initial condition.}\label{fig2}
\end{center}
\end{figure}

Now, we study the cosmic evolution of the KHDE in the presence of the interaction between two dark components. 
For this purpose we again solve Eq. (\ref{Hdot2}) numerically and then use it to plot the evolution of the cosmological parameters.
It should be noted that in this section, fixed values of $ H_0=67.9, \Omega_m=0.3, \kappa=0.8, \phi_0=0.4, n=0.06 $ and $ \omega=10 $ are considered.
We have plotted the behavior of dimensionless density parameter $ \Omega_D $,
the deceleration parameter $ q $ and the EoS parameter  with respect $ z $  in Fig. \ref{fig2} 

Using Eqs. {\ref{Omega}}  and (\ref{Hdot2}),
we can obtain the evolution of $\Omega_D(z)$ for interacting KHDE which has been plotted  in the upper diagram in Fig.~(\ref{fig2}).
From this figure, it is clear that~$ \Omega_D\rightarrow 0.2 $ at the early times of the universe ($ z\rightarrow \infty $), according to Eq.~(\ref{Friedeq03}) and the DE dominates ($ \Omega_D\rightarrow1 $) at late-time ($ z\rightarrow-1 $). \\

Behavior of the deceleration parameter $ q(z) $ is illustrated in the middle diagram of Fig. \ref{fig2}.
As it is observed, our universe undergoes a transition from deceleration to acceleration phase at the redshift value around
$ 0.3<z<0.6 $.  Such a transition for $ \alpha>0 $ and $ \beta<0 $ happens at a higher redshift (earlier universe). 

From the latest diagram one can easily see that the EoS parameter can remain in the quintessence regime
or can also  enter the phantom regime, which depends on the coupling  constants $ \alpha $ and $ \beta $.
 
\begin{figure}[htp]
\begin{center}
\includegraphics[width=8cm]{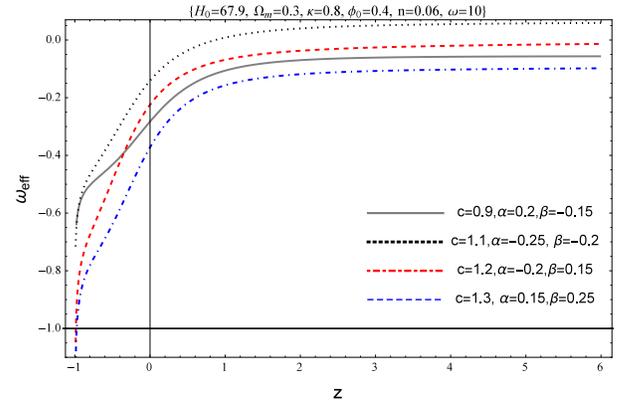}
\caption{The evolution of the effective EoS parameter $\omega_{\rm eff}$ versus redshift parameter $z$ for interacting THDE. }\label{figqQ}
\end{center}
\end{figure}
\section{Stability}

In this section we would like to study the classical stability of the KHDE model against small perturbations in the background.
In the perturbation theory, the sign of squared of the sound speed, $ v_s^2 $, determines the stability of the background evolution. 
For $ v_s^2>0 $ the model is stable against the perturbations, since the given perturbation propagates in the environment,
while for $ v_s^2<0 $ the amplitude of perturbations grows within the environment and consequently the model is unstable.
The squared sound speed $ v_s^2 $ is given by
\begin{equation}\label{v_s}
v_s^2=\frac{dp}{d\rho_D}=\frac{\dot{p}}{\dot{\rho}_D}.
\end{equation}
By differentiating $p_{D}$ with respect to time  together with inserting the result into Eq.~(\ref{v_s}), 
and using Eq. (\ref{rhodot}), we can get
\begin{equation}\label{v1}
v_s^2=\omega_D+\frac{\omega_D^\prime}{2\delta n+2(2-\delta)\frac{\dot{H}}{H^2}},
\end{equation}
for the squared sound speed.

\subsection{Non-interacting case}
\begin{figure}[htp]
\begin{center}
\includegraphics[width=8cm]{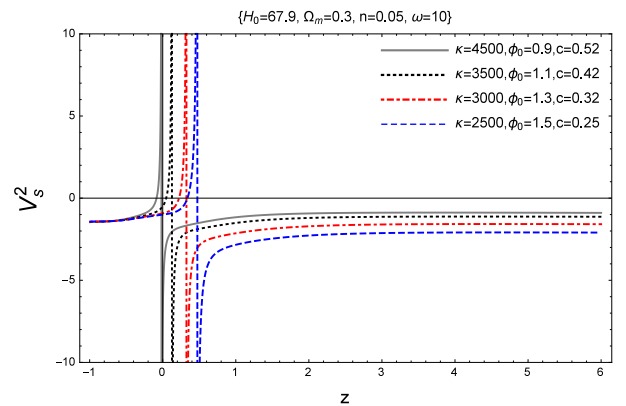}
\caption{The evolution of $ v_s^2$ against redshift parameter $ z $ for non-interacting KHDE
for he initial condition $ H_0=67.9, \Omega_m= 0.3, n=0.05, \omega=10$ and 
different values of $ \kappa, \phi_0, c $. }\label{figqV1}
\end{center}
\end{figure}

Taking the time derivative of Eq.~(\ref{EoS1}) and using Eqs.~(\ref{Omega1}),~(\ref{rhodot}),~(\ref{Hdot1}) and
(\ref{v1}), we can obtain $ v_s^2 $ for the non-interacting KHDE with the Hubble cutoff in BD cosmology.
Since this expression is too long, we shall not present it here, and only plot it in Fig. \ref{figqV1}.
From this figure we see that the non-interacting KHDE model in BD gravity is unstable $ (v_s^2<0) $.
\subsection{Interacting case}
\begin{figure}[htp]
\begin{center}
\includegraphics[width=8cm]{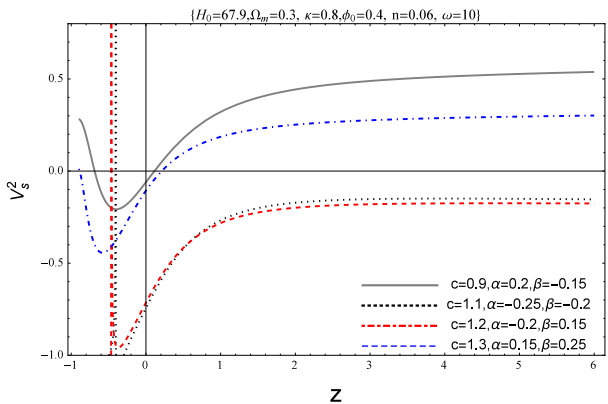}
\caption{The evolution of  $ v_s^2(z)$ for interacting KHDE in BD gravity.
The initial condition $ H_0=67.9, \Omega_m=0.3, n=0.05, \omega=10$ and 
	different values of $ \kappa, \phi_0, c $ are adopted. }\label{figV2}
\end{center}
\end{figure}

We now calculate the squared of the sound speed for the interacting KHDE model.
By taking the time derivative of Eq. (\ref{EoS2}), 
and combining the result with Eqs. (\ref{rho1}), (\ref{rhodot}), (\ref{Hdot2}) and (\ref{v1}), we can obtain $ v_s^2 $.
Again, since this expression is too long, we do not present it here.
We have plotted $ v_s^2 $ of the interacting KHDE in the BD gravity in Figs. \ref{figV2}.
As it is clear, in the presence of interaction term, the squared of the sound speed is positive 
in a period of time meaning that the interacting KHDE can be stable 
for $ \alpha>0 $ (whether $ \beta $ is positive or negative).

\section{Concluding remarks}
In this paper we studied the results of using the Kaniadakis entropy to construct a holographic dark energy model
(called Kaniadakis holographic DE) in  the framework of BD theory. We considered the Hubble horizon as the IR cutoff and 
studied the behavior of KHDE model for both the interacting and non-interacting cases.
To this end we obtained the equation governing the evolution of Hubble parameter and used its numerical solution 
 in order to study the evolution of the corresponding cosmological parameters.\\
From the  behavior of deceleration parameter $ q $, we found that,  unlike the standard HDE in the BD gravity, the KHDE with the Hubble cutoff in BD gravity can provide a setting for current accelerated expansion of the universe, even in the absence of an interaction between the two dark sectors of cosmos. In addition, the behavior of EoS parameter, for both interacting and non-interacting cases, shows that the KHDE model can cross the phantom divide depending on the values of free model parameters. Moreover, the EoS parameter in the presence of interaction shows an interesting behavior, i.e., it can represent a quintessence regime ( $\omega_D>-1 $), a phantom one
 $ (\omega_D<-1) $ or cosmological constant $ (\omega_D=-1)$. 
In order to study the fate of the universe filled with DM and KHDE, we have  
plotted the behavior of the effective EoS parameter $ \omega_{\rm eff}(z) $ for various interacting and non-interacting parameters in Figs. \ref{figweff1} and \ref{figqQ}. We observed that, at high redshifts, the effective matter content of the universe is in the dust form and evolves to a DE form at late times. This study indicated that as $\omega_{\rm eff}\rightarrow -1$ for $z\rightarrow-1$, then the
occurrence of a Big-Rip singularity at a certain time in the future is possible.\\
Finally, the analysis of the squared of sound speed $ v_s^2 $ revealed that non-interacting KHDE in BD gravity is
unstable against small perturbations in the background.  This implies that considering an interaction between 
two dark sectors seems more reasonable. We therefore found that interacting KHDE can provide stability in the period of time.


\end{document}